\documentclass[]{iopart}

\usepackage{graphicx}
\usepackage{dcolumn}
\usepackage{bm}

\begin{document}

\title{Tunable charge carriers and thermoelectricity of single-crystal Ba$_8$Ga$_{16}$Sn$_{30}$}

\author{M A Avila$^1$, D Huo$^{1,2}$, T Sakata$^1$, K Suekuni$^1$, T Takabatake$^1$}
\address{%
$^1$ Department of Quantum Matter, ADSM, Hiroshima University,
Higashi-Hiroshima 739-8530 Japan
}%
\address{%
$^2$ Department of Applied Physics, Hangzhou Dianzi University,
Hangzhou 310018, China
}%

\begin{abstract}
We have grown single crystals of the type-VIII intermetallic
clathrate Ba$_8$Ga$_{16}$Sn$_{30}$ from both Sn and Ga flux,
evaluated their compositions through electron microprobe analysis
and studied their transport properties through measurements on
temperature dependent resistivity, thermopower and Hall
coefficient. Crystals grown in Sn flux show n-type carriers and
those from Ga flux show p-type carriers, whereas all measured
compositions remain very close to the stoichiometric 8:16:30
proportion of Ba:Ga:Sn, expected from charge-balance principles.
Our results indicate a very high sensitivity of the charge carrier
nature and density with respect to the growth conditions, leading
to relevant differences in transport properties which point to the
importance of tuning this material for optimal thermoelectric
performance.
\end{abstract}



\section{Introduction}

Research on intermetallic clathrate compounds with general formula
A$_8$X'$_{16}$X$_{30}$ (A = Ba, Sr, Eu; X' = Al, Ga, In; X = Si,
Ge, Sn) has increased significantly over the past 10 years, since
the proposal that the generic class of caged compounds with guest
atoms might be good candidates to fulfill the phonon-glass
electron-crystal (PGEC) concept of a thermoelectric material with
potential for applications \cite{sla95a}. The cages in these
clathrates are formed by X'X host atoms connected through
diamond-like hybridized orbitals, while the A guest atom lies
within the cage and donates two electrons to it. For each divalent
A ion in this particular family, the replacement of two X atoms
with X' ions in the cage appears to be the required charge-balance
condition for these two electrons to be accepted and for the
structure to be stabilized.

The potential phonon-glass behavior results from the fact that the
A$^{2+}$ ion can in certain cases be loosely bound inside a cage
that is slightly oversized in comparison with its ionic radius,
and is thus able to achieve a rather independent ``rattling''
motion that tends to scatter low-frequency acoustic phonons,
responsible for a good fraction of the heat conduction in a
crystalline lattice \cite{cohn99a}. In the case of
Sr$_8$Ga$_{16}$Ge$_{30}$ and Eu$_8$Ga$_{16}$Ge$_{30}$, the
additional presence of a four-fold splitting of an A atom site and
the consequent tunnelling of this ion between these four
off-center sites is claimed to be the source of truly glass-like
behavior in their thermal conductivities at low temperatures
\cite{sal01a,cha01a,kep02a,zer04a}. The electron-crystal
characteristic of these clathrates is preserved because the charge
carriers remain within the cage network and are not significantly
affected by the rattling motion of the guest atom.

We have recently reported a thorough characterization of the basic
physical properties of the Type-VIII clathrate
Ba$_8$Ga$_{16}$Sn$_{30}$, grown as large single crystals out of Sn
flux \cite{huo05a}. Electron probe microanalysis (EPMA) showed
that these crystals grew with composition very close to 8:16:30
despite the excess Sn flux, confirming the relevance of the above
mentioned charge balance in the stabilization of the structure.
These crystals showed behavior consistent with a heavily-doped
$n$-type semiconductor with large negative thermopower and
negative Hall coefficient. They also showed low lattice thermal
conductivity of order 1~W/m~K resultant from the rattling of Ba
ions, which was evidenced by large isotropic displacement
parameters in single crystal x-ray diffraction analysis and the
presence of Einstein vibrational degrees of freedom in heat
capacity analysis.

Towards the end of the experimental investigations for that work,
we also succeeded in growing single crystalline
Ba$_8$Ga$_{16}$Sn$_{30}$ out of Ga flux, and EPMA analysis once
again showed a crystal composition very close to 8:16:30. Since
these Ga-flux crystals did not grow as easily or as large as the
Sn-flux ones, and the final composition was not significantly
changed, they were not investigated in detail for that initial
characterization. However, a recent study of sintered samples of
Ba$_8$Ga$_{16+x}$Sn$_{30-x}$ ($x$ being the \emph{nominal}
composition) showed that the carrier changes from electron-type
for $x\leq-2$ to hole-type for $x\geq-1$ \cite{kur00a}. This
result invited a more careful investigation on the dependence of
our crystal's properties with their growth conditions.

In the present work, we extend and complete the basic
characterization of single crystalline Ba$_8$Ga$_{16}$Sn$_{30}$
with a comparative study between Sn-flux and Ga-flux grown
crystals. We will show that the transport properties are in fact
highly sensitive to composition and/or growth conditions, and thus
this material has a potential \emph{tunability} that may in
principle allow significant improvements in terms of its
thermoelectric performance, in conjunction with other materials
science techniques for such.

\section{Experimental Details}

Four single crystal batches were prepared and measured for this
work: two using excess Sn flux and two using excess Ga flux.
Samples from these batches will be referred to as Sn\#1, Sn\#2,
Ga\#1 and Ga\#2 respectively. Sn\#1 samples are from the same
crystal growth described in detail in the previous work
\cite{huo05a} using a Ba:Ga:Sn starting proportion of 8:16:60, and
Sn\#2 samples are from a second batch of crystals grown
essentially by the same route. With Sn flux the crystals tend to
grow very large (limited mostly by the total mass of the starting
reagents and by the quartz tube size of order 10~mm in diameter)
and have very well defined polyhedral surface facets.

For the Ga\#1 and Ga\#2 growths the starting elemental proportions
were 8:26:30 and 8:50:30 respectively. The high-purity elements
were sealed in evacuated quartz tubes, soaked above
1150~$^{\circ}$C for 2-3 hours, cooled over 10 hours to
550~$^{\circ}$C and then slowly cooled over 100 hours to
420~$^{\circ}$C. At this point the ampoules are quickly removed
from the furnace and the remaining molten Ga flux separated by
centrifuging. If the crystals did not grow to sufficient size in
this first round, the solidified material was resealed with enough
new Ga to complete the initial Ba:Ga:Sn proportion once again, and
only the second part of the ramp was repeated. Polyhedral crystals
grown from this approach were never larger than about 3 mm in
diameter, and usually much smaller. Some of our growth attempts
also resulted in large plate-like crystals rather than the
expected polyhedrons. EPMA analysis showed that these plates are a
new compound BaGa$_3$Sn, which we intend to characterize and
describe in a future communication.

To ensure the best possible reliability of a comparative EPMA
evaluation of the Sn- and Ga-flux grown Ba$_8$Ga$_{16}$Sn$_{30}$
crystals, we mounted one crystal from each batch on a single
sample holder and measured all four in a single experimental
session at a JEOL JXA-8200 microanalyzer, using the same reference
materials for Ba, Ga and Sn. The results (averaged over 5
different regions for each crystal) are shown in table~\ref{EPMA}.
All crystal composition remain close to the nominal 8:16:30
proportion, but there are clear distinctions. Sample Sn\#1 is,
within experimental error, in the ideal stoichiometry. Samples
Sn\#2 and Ga\#1 show an average slight excess of Sn and Ga
respectively, and sample Ga\#2 showed the greatest
off-stoichiometry composition (excess Ga).

\begin{table}
\caption{\label{EPMA} Relative Ba:Ga:Sn content in the four
measured crystals as determined by electron-probe microanalysis,
and Fermi energies derived from thermopower analysis.}
\begin{center}
\begin{tabular}{lccrcc}
\hline
Batch & Starting & \multicolumn{3}{l}{Crystal Composition} & E$_F$\\
Name & Composition & Ba & Ga & Sn & (meV)\\
\hline
Sn\#1 & 8:16:60 & 8 & 16.0 & 30.0 & 88\\
Sn\#2 & 8:16:60 & 8 &  15.9 & 30.1 & 175\\
Ga\#1 & 8:26:30 & 8 &  16.1 & 29.9 & 225\\
Ga\#2 & 8:50:30 & 8 &  16.3 & 29.7 & 525\\
\hline
\end{tabular}
\end{center}
\end{table}

DC electrical resistivity experiments in the range of 4-300~K were
performed on a home-made probe inserted into a glass cryostat.
Crystals were shaped into elongated bars using a spark cutter and
gold wire contacts were attached with silver paste, in standard
4-probe geometry. For each point the average voltage was obtained
from measurements with the current (1-10 mA) applied in opposite
directions. Thermopower from 4-300~K was measured by a
differential method, performed on a home-made probe inserted into
a glass cryostat. The bar-shaped samples were suspended between
two electrically isolated Copper blocks, across which a
temperature difference of 0.05-0.3~K was applied. Hall coefficient
from 4-300~K was measured on a home-made probe by a DC technique
in a field of 1~T applied by a conduction magnet. Crystals were
shaped into a thin rectangular plate using a spark cutter and gold
wire contacts were attached by spot welding.

\begin{figure}[htb]
\begin{center}
\includegraphics[angle=0,width=106mm]{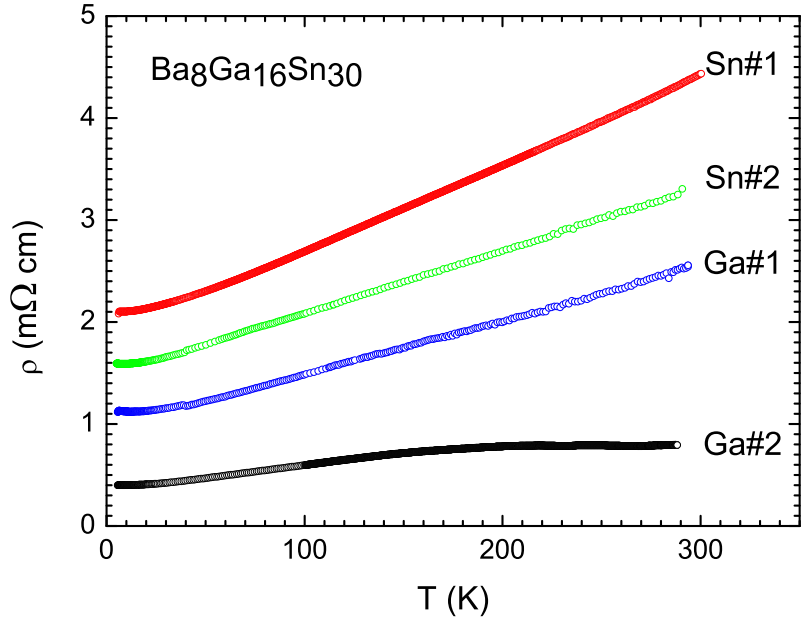}
\caption{\label{res} (color online) Temperature dependence of the
electrical resistivity $\rho(T)$ of Ba$_8$Ga$_{16}$Sn$_{30}$
crystals grown from Sn and Ga flux.}
\end{center}
\end{figure}

\begin{figure}[htb]
\begin{center}
\includegraphics[angle=0,width=106mm]{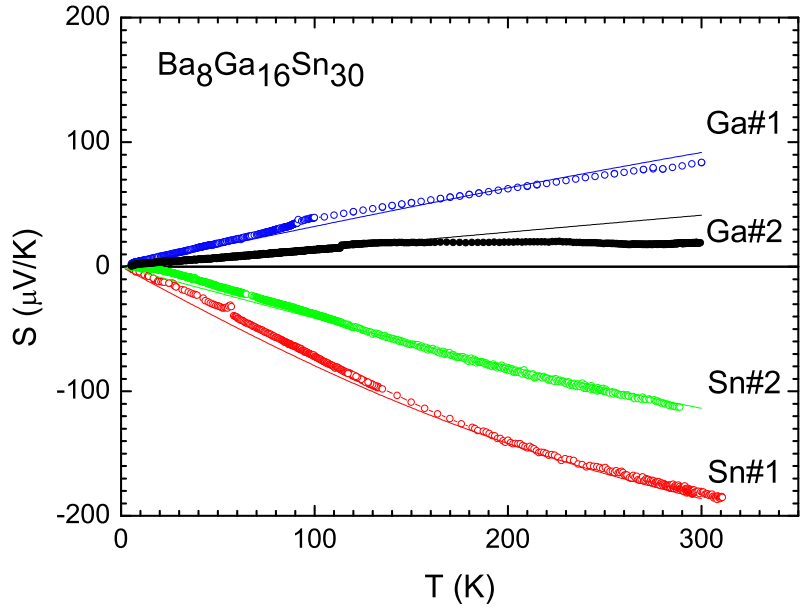}
\caption{\label{thermo} (color online) Temperature dependence of
the thermoelectric power $S(T)$ of Ba$_8$Ga$_{16}$Sn$_{30}$
crystals grown from Sn and Ga flux. Solid lines are the best fits
of a single-band model.}
\end{center}
\end{figure}

\section{Experimental Results}

The resistivity behavior $\rho(T)$ below room temperature of all
samples is shown in Fig.~\ref{res}. They are in the m$\Omega$~cm
range and show monotonically decreasing values upon cooling,
characteristic of heavily-doped semiconductors or low
carrier-density metals. Since these are large, dense and carefully
cut samples, (rectangular cross-sections of order $A=1$~mm$^2$)
the uncertainty in estimating the geometrical factor $A/d$ is
relatively small (less than 5\%), so the differences observed in
the calculated resistivities should not be resultant from trivial
geometrical uncertainty. The quasi-stoichiometric Sn\#1 sample has
the highest resistivity, consistent with the idea that the ideal
material tends towards a narrow-gap semiconductor. But none of the
Ba$_8$Ga$_{16}$Sn$_{30}$ crystals grown so far have shown
semi-conducting resistivity like that of Ba$_8$Ga$_{16}$Ge$_{30}$
\cite{umeo05a}, so the question remains open whether there is a
true gap or a pseudo-gap in the ideal material. The Sn\#2 and
Ga\#1 samples have smaller resistivities, as one would expect from
the electron and hole doping effect of their respective
off-stoichiometries, and the Ga\#2 sample has the lowest
resistivity, consistent with its largest doping level.

The differences between the type of charge carriers and its
densities become much more evident in the thermopower
measurements. In Fig.~\ref{thermo} we show $S(T)$ for the same
samples in Fig.~\ref{res}. As shown in the previous work, Sn\#1
sample has negative, monotonically decreasing thermopower in the
entire temperature interval (n-type carriers) and reaches a
relatively large magnitude of $-185\mu$V/K at 290~K. The Sn\#2
sample shows somewhat smaller but qualitatively similar behavior,
reaching $-113\mu$V/K at 290~K. In contrast, the two Ga flux
samples show positive $S(T)$ over the entire measured interval,
indicating p-type carriers. Sample Ga\#1 reaches $+83\mu$V/K at
290~K, while sample Ga\#1 remains below $+20\mu$V/K. The solid
lines show the best fits of these $S(T)$ curves to the equation

\begin{equation}
S(T)=\frac{k_B}{e}\left(\frac{4F_3(E_F/k_B T)}{3F_2(E_F/k_B T)}-
E_F/k_B T\right)
\end{equation}

derived for a single parabolic band model \cite{lov77a}, where
$F_n$ is the Fermi-Dirac integral of order $n$, and the only
fitting parameter is the Fermi energy $E_F$, whose respective
values are listed in table~\ref{EPMA}. It is reasonable to assume
that this system should be better described by a two-band model
though.

\begin{figure}[htb]
\begin{center}
\includegraphics[angle=0,width=102mm]{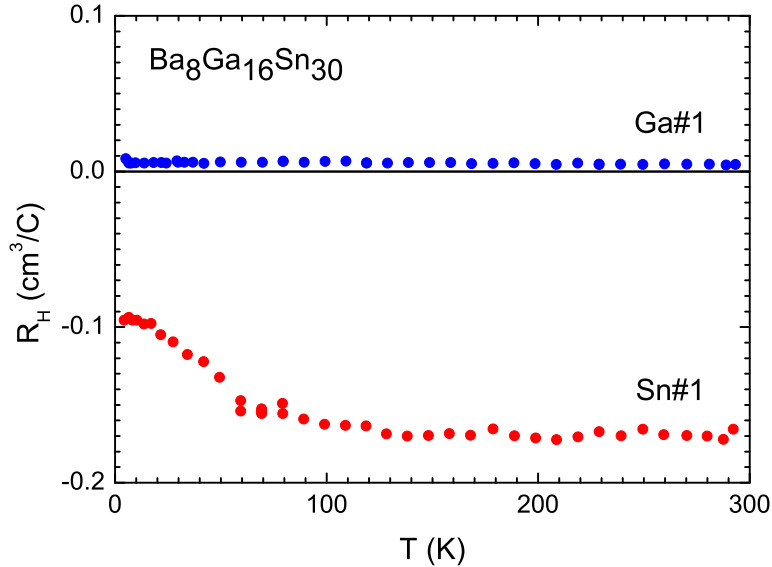}
\caption{\label{hallcoef} (color online) Temperature dependence of
the Hall coefficient $R_H$ of Ba$_8$Ga$_{16}$Sn$_{30}$ crystals
grown from Sn and Ga flux.}
\end{center}
\end{figure}

\begin{figure}[htb]
\begin{center}
\includegraphics[angle=0,width=102mm]{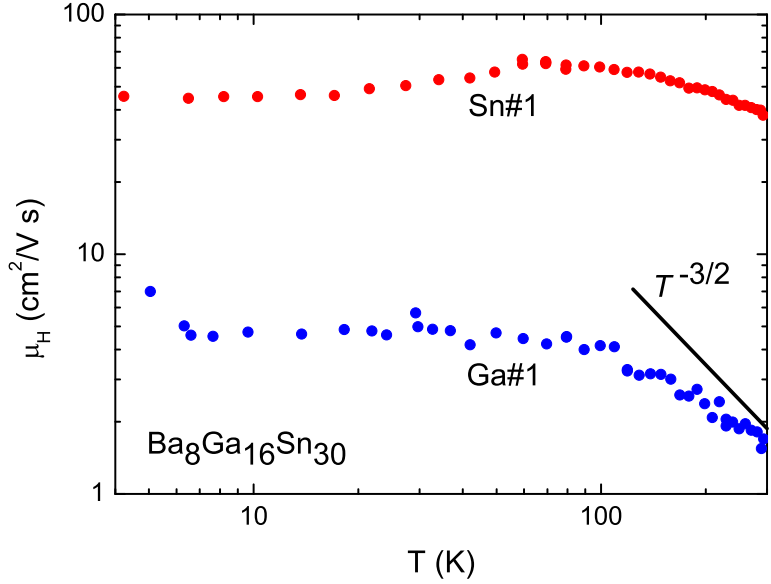}
\caption{\label{hallmob} (color online) Temperature dependence of
the Hall mobility $\mu_H$ of Ba$_8$Ga$_{16}$Sn$_{30}$ crystals
grown from Sn and Ga flux.}
\end{center}
\end{figure}

Further confirmation of these distinctions for different fluxes
came with the Hall coefficient measurements, shown in
fig.~\ref{hallcoef}. A sample from batch Sn\#1 shows negative
$R_H$ in the entire temperature interval, while a sample from
batch Ga\#1 showed a very small and positive $R_H$. From the
single-band model \cite{lov77a}, the carrier concentration
$n=1/eR_H$ obtained from these measurements at 300~K are
$3.7\times10^{19}$~electrons/cm$^3$ and
$1.3\times10^{21}$~holes/cm$^3$ respectively. The temperature
dependence of the Hall mobility $\mu_H(T)=|R_H(T)|/\rho(T)$
derived from these measurements is plotted in Fig.~\ref{hallmob}.
The respective values at 290~K are 39 cm$^2$/V~s and 1.8
cm$^2$/V~s, and in both samples a negative slope is seen at higher
temperatures, approaching a $T^{-3/2}$ law expected from acoustic
phonon scattering \cite{seeg85a}, indicating that this should be
the dominant scattering mechanism in this region.

\section{Discussion}

The principle of operation of a thermoelectric cooling device
requires a pair of materials, one n-type and one p-type, coupled
electrically in series and thermally in parallel \cite{nol01a}.
The device's performance optimization involves not only the
properties of each individual branch, but a combined optimization
of the thermoelectric pair. Thus, an important goal of research in
this area is to not only find good thermoelectric materials of
both carrier types, but also find materials that may be fine-tuned
with respect to their figure of merit $ZT=S^2T/\rho\kappa$. Our
experiments show that Ba$_8$Ga$_{16}$Sn$_{30}$ samples may be a
candidate to fulfill most of these requirements, if its present
$ZT$ of order 0.15 at room temperature \cite{huo05a} can be
improved closer to unity.

The existence of n-type and p-type samples has already been
observed in sintered samples with nominal compositions
Ba$_8$Ga$_{16+x}$Sn$_{30-x}$ \cite{kur00a}. It was reported that a
gradual decrease of n-type carriers for $-3<x<-2$ is followed by a
gradual increase of p-type carriers for $-1<x<+2$, with a possible
``quasi-insulating'' region around $x=-1.5$. Our results clearly
show that the flux grown crystals are strikingly more sensitive in
terms of sample composition, such that off-stoichiometries of
$x=\pm0.1$ from the ideal value are enough to cause the drastic
change. This also raises the question of whether the largely
deviated \emph{nominal} compositions assumed in the polycrystals
truly reflect their actual intra-grain clathrate compositions.

If we can understand the reasons behind the observed differences
in carriers and transport properties, we may be able to control
these through the growth parameters and/or through post growth
thermal treatment. The first issue is whether the observed
differences are indeed primarily associated with
off-stoichiometry. Our experimental data give good evidence that
there is in fact a correlation with the Ga-Sn proportion, such
that the transport properties can be directly related to the ionic
self-doping level.

But the transport properties may not be directly related only to
differences in composition. The next logical step is to account
for differences in the distribution of Ga-Sn atoms throughout the
cage structure. Contrary to the Ga-Ge clathrates where, due to
size similarity of these two atoms, they show mostly random
distribution throughout all three crystallographic sites of the
type-I clathrate structure (although there are some indications of
possible preferential occupations) \cite{cha01a,cha00a,zha02a} in
Ba$_8$Ga$_{16}$Sn$_{30}$ the size difference between Ga and Sn
ions is large enough that there is a clear preferential occupation
of at least three of their four different sites in the type-VIII
clathrate structure. Assuming that random occupation of a given
site implies a proportion close to Ga$_{16}$Sn$_{30}$ in that
site, the single crystal XRD refinement on the Sn\#2 crystal
performed for the previous work \cite{huo05a} revealed that Sn
tends to favor the X(1) site (Ga$_{8.5}$Sn$_{37.5}$) and X(2) site
(Ga$_{7.3}$Sn$_{38.7}$) while Ga tends to occupy the X(4) site
(Ga$_{35.3}$Sn$_{10.7}$) which has the shortest bond distances
between neighbors, and the X(3) site remains more randomly
occupied by both atoms (Ga$_{14.4}$Sn$_{31.6}$). We performed a
preliminary comparative investigation of the Sn\#2 crystal before
and after annealing for one week at 480~$^{\circ}$C, which showed
only a 12\% decrease in its thermopower above room temperature, so
that either the annealing condition was ineffective in
significantly rearranging the as grown Ga-Sn distribution, or the
differences in distribution may be indeed secondary with respect
to thermoelectric properties, in comparison to composition
differences.

\section{Conclusion}

Single crystals of Ba$_8$Ga$_{16}$Sn$_{30}$ grown from Sn flux and
Ga flux have shown significantly different charge carrier
densities and transport properties, which we could correlate to
the relative Ga-Sn content in the crystals. The crossover between
n-type and p-type carriers occurs within a very narrow range
around the ideal 8:16:30 stoichiometry, in agreement with
charge-balance principles and contrary to previously reported
preliminary studies on sintered polycrystals. Such characteristics
point to the possibility of tuning the charge carrier nature and
density, as well as the transport properties, through the sample
preparation process and possibly through post growth annealings, a
desirable feature in any material intended for thermoelectric
applications. Our next step is now to investigate the
low-temperature heat capacity/conductance behavior of the
carrier-tuned Ba$_8$Ga$_{16}$Sn$_{30}$ crystals, in search of
further information on the vibrational behavior of the Ba guests
ions under different host cage environments.

To our satisfaction, almost at the same time we first submitted
this manuscript the research group from Dresden published two
excellent works \cite{pach05a,ben05a} describing how their
polycrystalline Eu$_8$Ga$_{16}$Ge$_{30}$ samples can be tuned with
respect to composition, carrier concentration and thermoelectric
properties, although in their case only n-type samples were found.
Their results also challenge the current ``tunneling'' models for
glass-like thermal conductivity (which we mentioned in the
introduction section) attributing this low-temperature behavior to
phonon-charge-carrier scattering instead.

\ack

We thank Y Shibata for the electron-probe microanalysis. This work
was financially supported by the COE Research (CE13CE2002) and the
priority area ``Skutterudite'' (No. 15072205) through
Grants-in-Aid from MEXT, Japan. D Huo is also supported by
National Science Foundation of China (No.50471008).

\end{document}